\documentclass{PoS}
\usepackage{cite}

\title{High density matter physics at J-PARC-HI}

\ShortTitle{High density matter physics at J-PARC-HI}

\author{\speaker{Takao Sakaguchi}, for the J-PARC-HI collaboration\\
        Brookhaven National Laboratory, Physics Department\\
        E-mail: \email{takao@bnl.gov}}

\abstract{Physics prospective of the high density matter using heavy ions
collisions is presented. The J-PARC-HI project which is a unique lab
to tackle the high density matter physics is described. The world highest
rate of heavy ion beam of 10$^{11}$\,Hz is aimed at J-PARC-HI
which enables us to perform measurements of hadrons, 
fluctuation of conserved quantities, dileptons,
multi-strange hypernuclei. New event selections are also
discussed to reach a highest baryon density on the ground.}

\FullConference{Corfu Summer Institute 2018 "School and Workshops on Elementary Particle Physics and Gravity"\\
		(CORFU2018)\\
		31 August - 28 September, 2018\\
		Corfu, Greece}

\begin{document}

\section{Introduction}
\begin{figure}[htbp]
    \centering
    \includegraphics[width=0.75\linewidth]{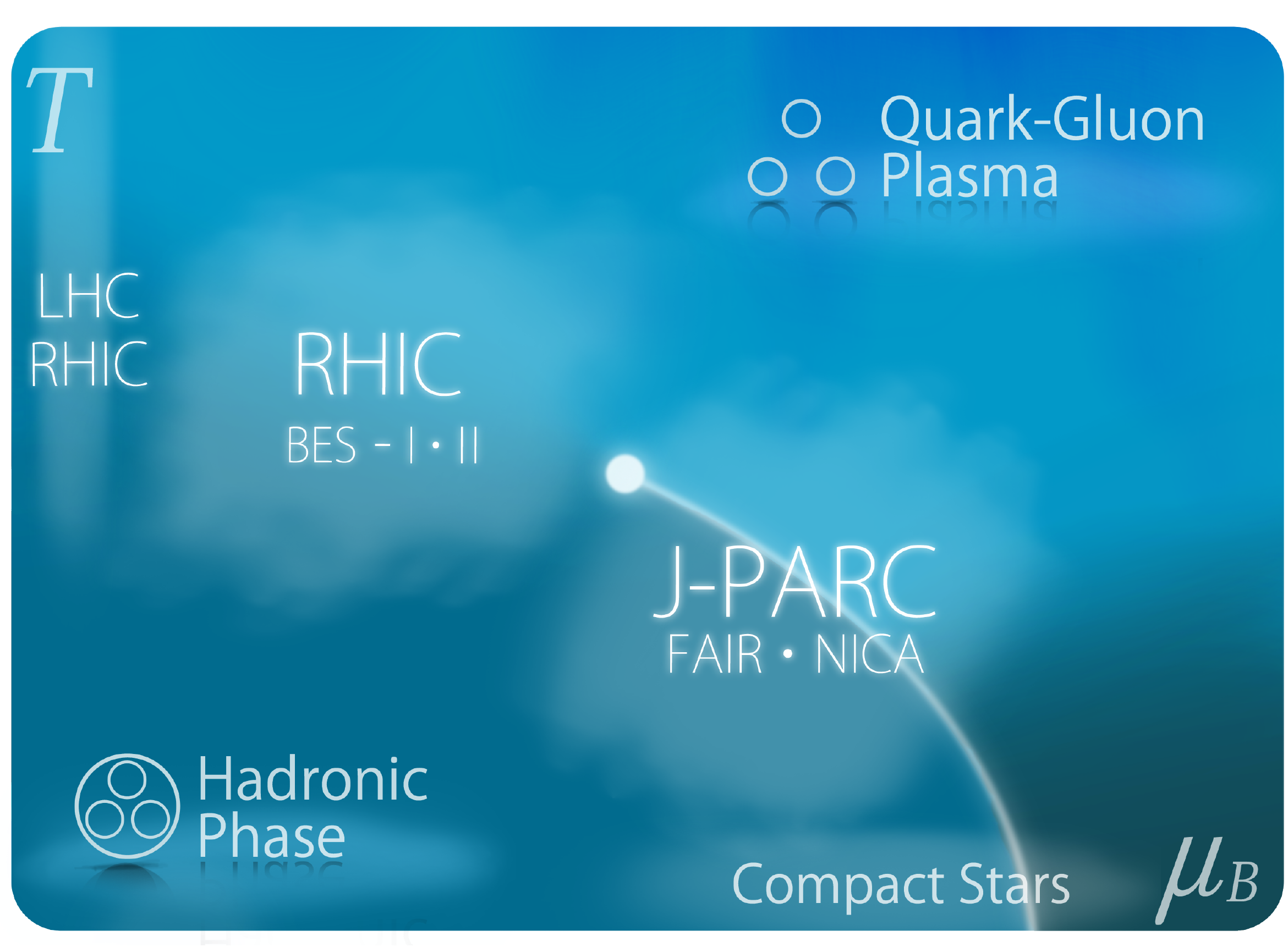}
    \caption{\label{fig1} QCD phase structure with the $T-\mu_b$ regions that existing and
    proposed experiments are covering.}
\end{figure}

Decades of the QCD matter study at high temperatures became fruition at RHIC and LHC
energies, where a new phase of the QCD matter, quark gluon plasma (QGP), was discovered
and has been studied in detail. However, the QCD has rich phase structure not
only in the high temperature side but also in the high baryon density side,
about which we know very little. Figure ~\ref{fig1} shows the QCD structure
known to date.
The high baryon density side supposedly contains the critical point where
the cross-over phase boundary turns into the 
first order phase transition. As going to even higher density, the matter phase
called quarkyonic or color superconductivity are expected to manifest.
This high density region is important not only for the nuclear physics,
but also for astrophysics, since these matters may be produced by
neutron-star mergers and/or at the core of neutron-stars.
For studying the properties
of such a high density matter, several future projects including
FAIR~\cite{fair}, NICA~\cite{nica}, and SPS have been proposed to
accelerate and collide heavy ions. The J-PARC is also planning to
accelerate heavy ions at the similar energies, but at a higher
beam intensity, after a modest upgrade to the already operating facility.
We will present perspective of physics outcome at J-PARC with the
heavy ion beams as well as the status of the project. This new project is
called as the J-PARC-HI project.

\section{Heavy ion acceleration at J-PARC}
\begin{figure}[htbp]
    \centering
    \includegraphics[width=1.0\linewidth]{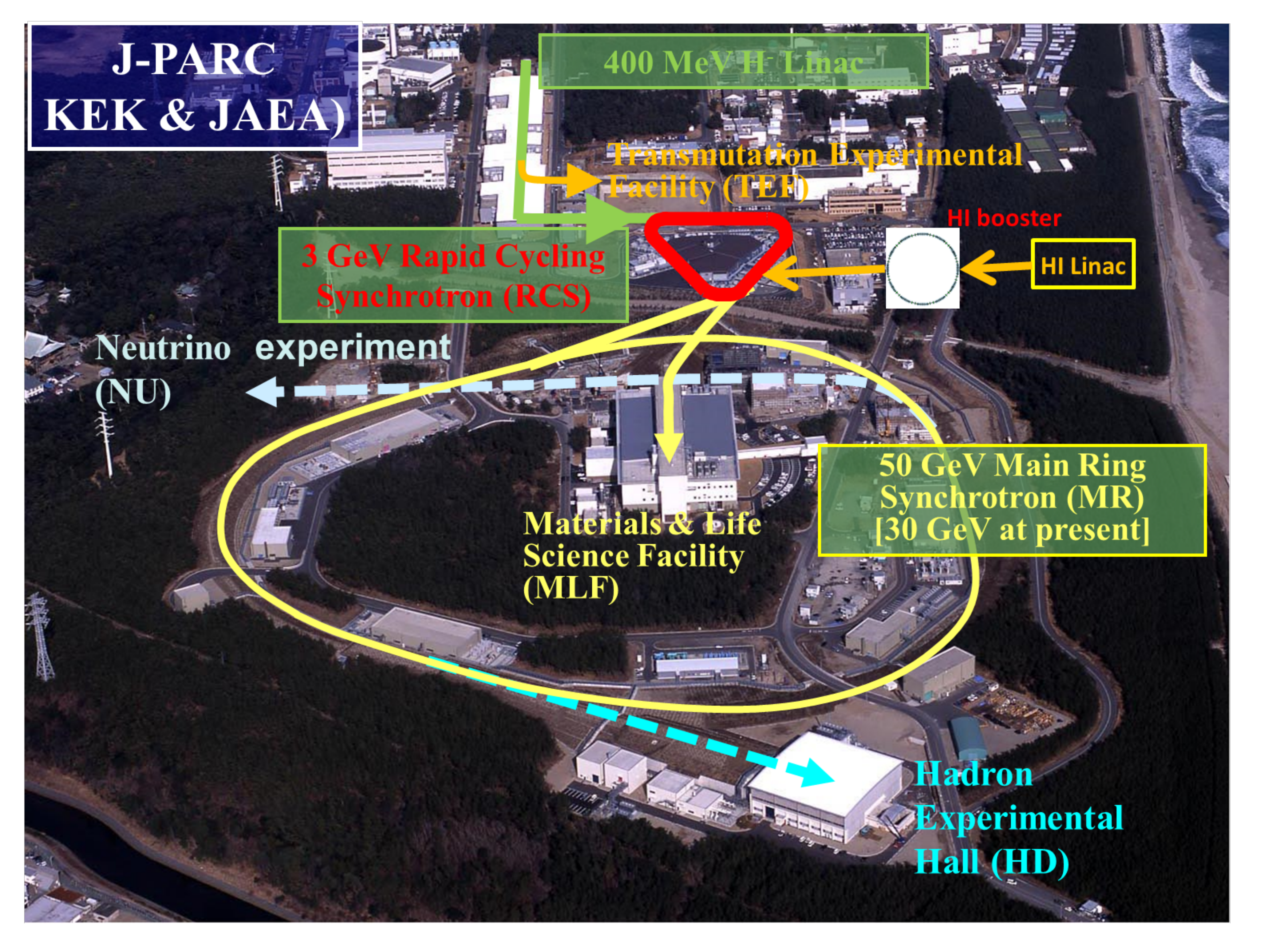}
    \caption{\label{fig21} Site view of the J-PARC facility with the illustration
    of the current configuration and the proposed upgrade for heavy ion acceleration.}
\end{figure}

\begin{figure}[htbp]
    \centering
    \includegraphics[width=1.0\linewidth]{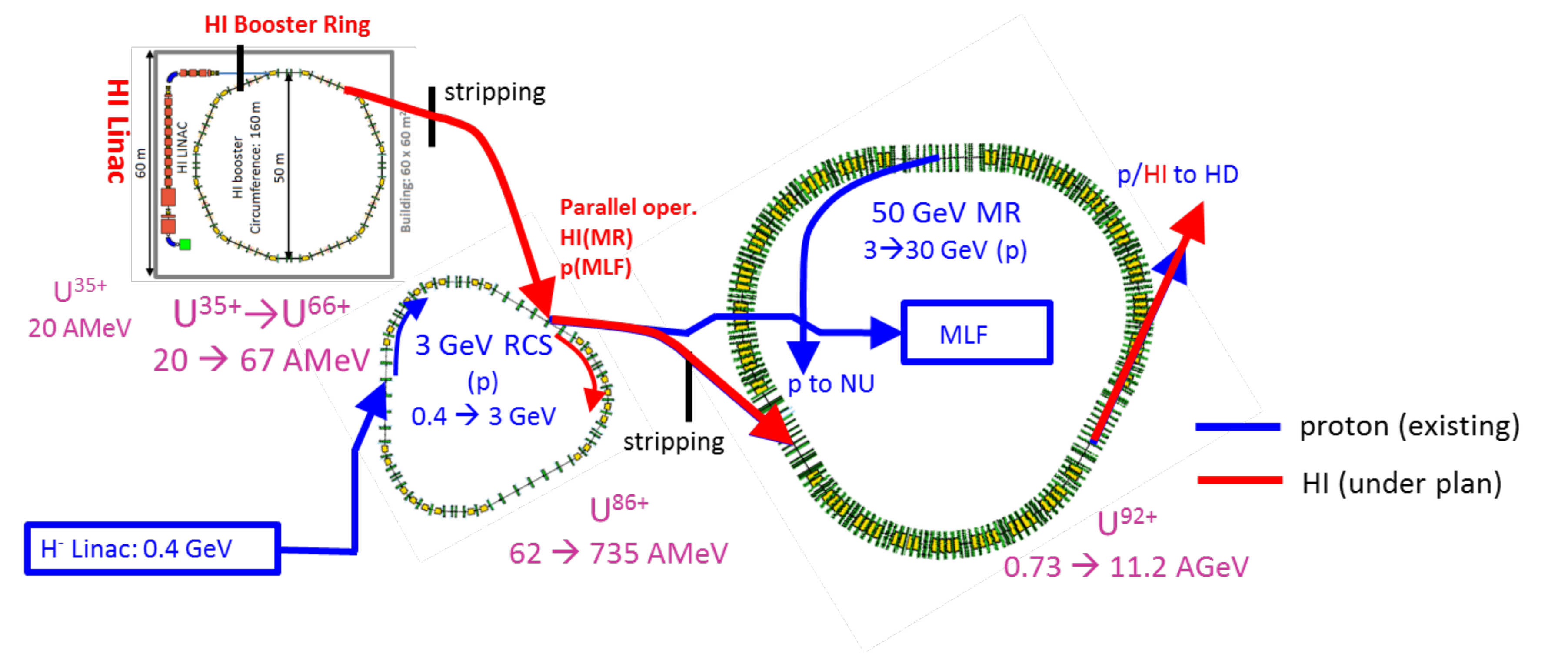}
    \caption{\label{fig22} Schematic presentation of the beam acceleration and transport scheme for the J-PARC-HI. During the time that protons are transported to the MLF
    (material and life science experimental facility), heavy ions can be
    accelerated and transported to the hadron experimental hall through the MR.}
\end{figure}

\begin{figure}[htbp]
    \centering
    \includegraphics[width=0.8\linewidth]{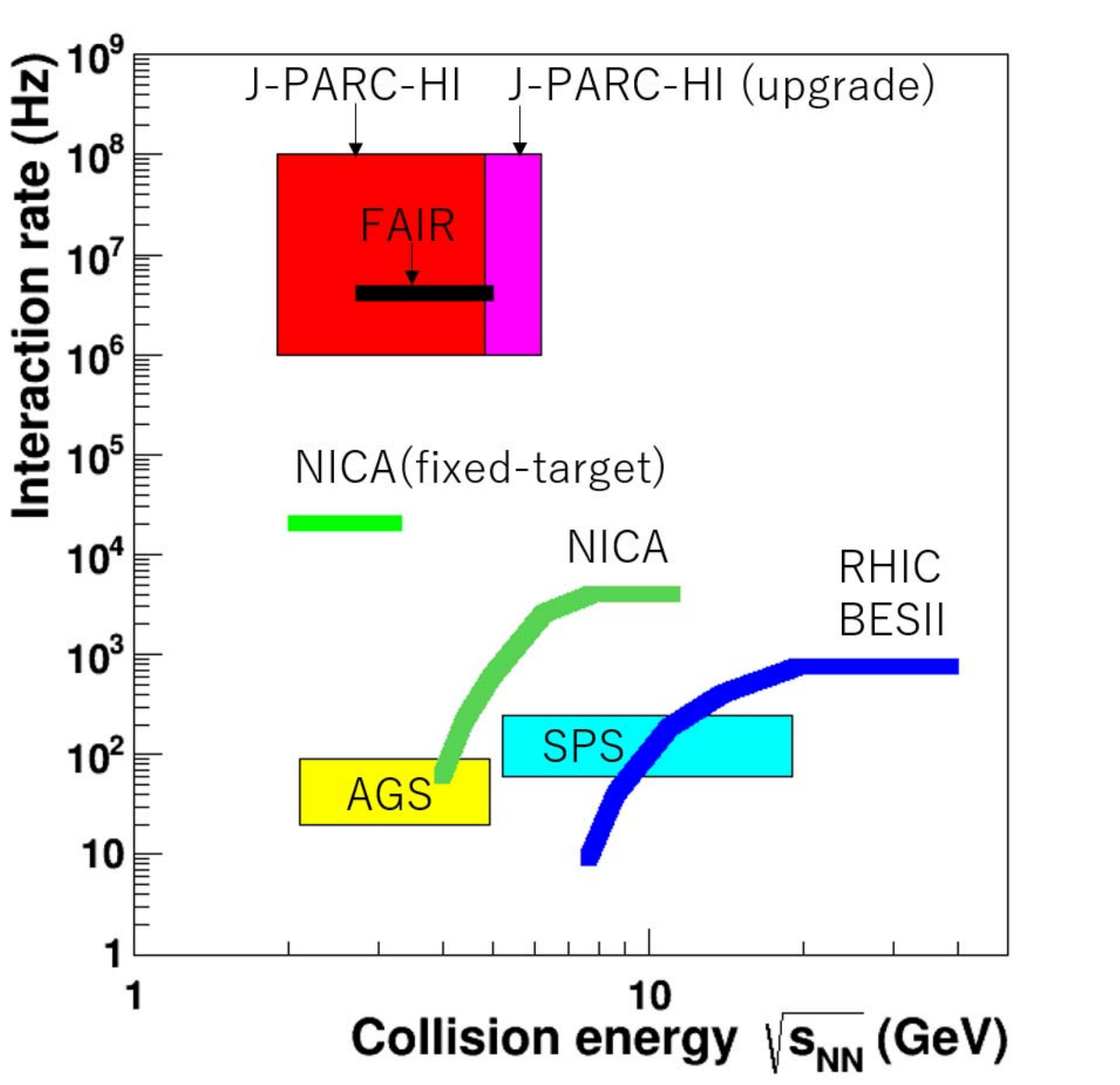}
    \caption{\label{fig23} Interaction rates and collision energies available at the various accelerator facilities in the world. FAIR and NICA (fixed target) provides similar beam energies as J-PARC-HI. The anticipated interaction rate can be higher for J-PARC-HI.}
\end{figure}

Figure~\ref{fig21} shows a site view of the J-PARC facility with the planned
addition for accelerating heavy ions.
The protons from the ion source are accelerated to 400\,MeV by the LINAC,
and injected to the 3\,GeV RCS (Rapid Cycle Syncrotron).
The MR (Main Ring) receives protons from the RCS and accelerates them
up to 50\,GeV (currently, it is operated up to 30\,GeV). For the machine
time for the material and life science facility (MLF), the protons are
provided directly from the RCS.
When accelerating heavy ions, one needs to construct a new ion source
followed by a new LINAC and a booster for injecting heavy ions to the RCS,
since the existing LINAC can only accept protons due to its rigidity constraint.
This acceleration scheme is depicted in Fig.~\ref{fig22} in more detail.
By tuning the operation parameters, this scheme can simultaneously accelerate
and provide protons and heavy ions, if the protons are to be provided to the
MLF and the heavy ions are to the hadron experimental hall. This will allocate
decent amount of machine time for heavy ion experiments since the current
machine time is dominated
by the experiments at the MLF. The upgraded J-PARC, J-PARC-HI, will be
flexible enough to provide
ions from protons up to Uranium at the lab energy of $E$=1--50\,GeV/n.
In case of Uranium, the incident energy will become 1--19\,GeV/n which
corresponds to $\sqrt{s_{NN}}$=1.9--6.2\,GeV. The beam rate will be as
high as 10$^{11}$\,Hz which is about one order higher than the
anticipated rate at FAIR/CBM. The comparison of the performance
with other facilities is shown in Fig.~\ref{fig23}.
With 0.1\,\% $\lambda_I$ target, one can expect the event rate as
high as
100\,MHz. This event rate will provide the one-year-run statistics at
AGS at BNL within five minutes of running at J-PARC-HI. The yields of
$\rho$,$\omega$, and $\phi$ decaying into electrons are expected to
be 10$^{10}$ to 10$^{12}$, that of hypernuclei will be 10$^{4}$ to
10$^{12}$~\cite{Andronic:2010qu}, and that of strangelets will be
1 to 10$^{2}$~\cite{BraunMunzinger:1994iq}, with a one month of
machine time at this event rate, without consideration of duty factors.

\section{Physics perspective}
\subsection{Collision dynamics at J-PARC-HI energy}
\begin{figure}[htbp]
\begin{minipage}{0.58\linewidth}
    \centering
    \includegraphics[width=1.0\linewidth]{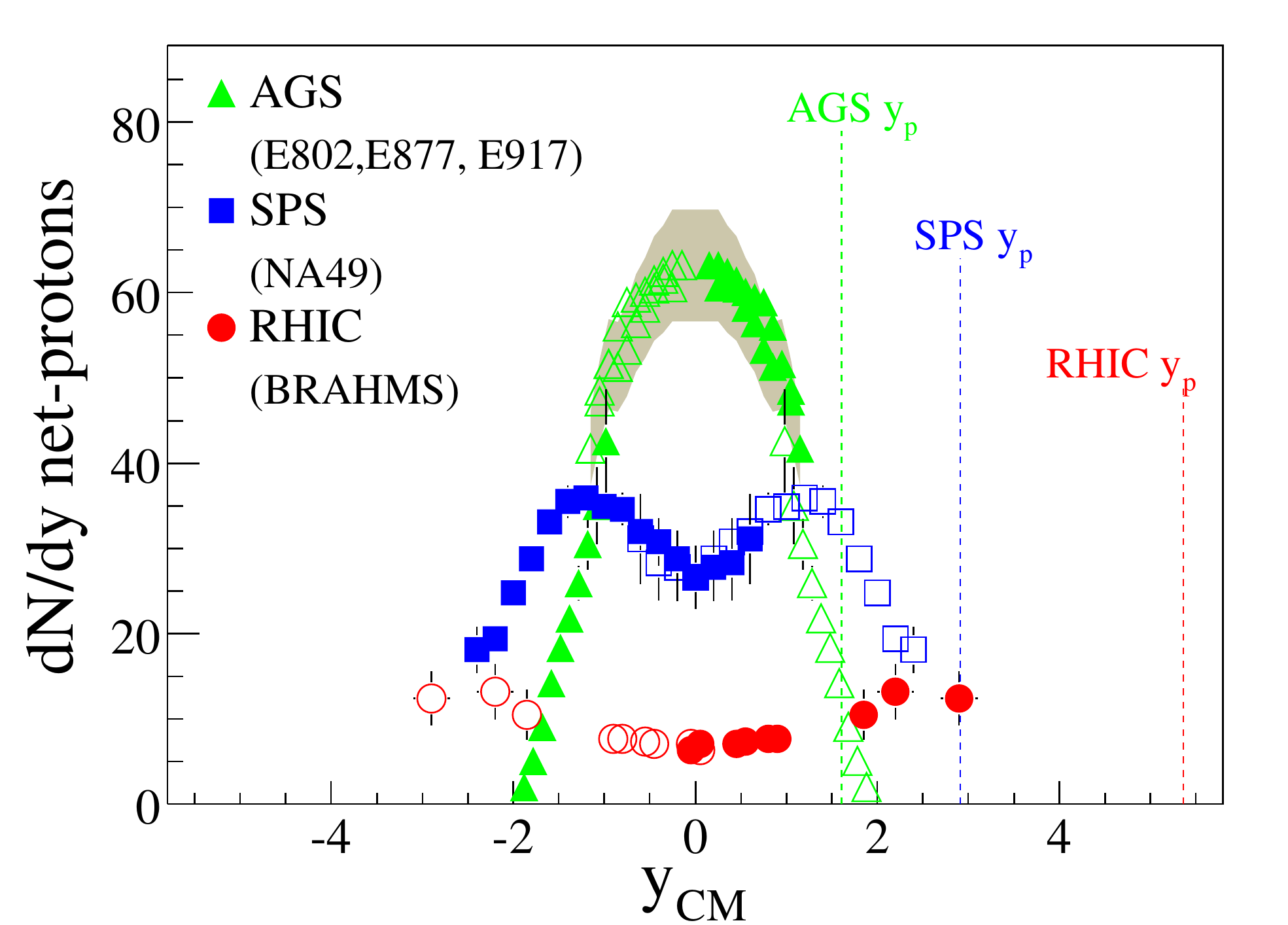}
\end{minipage}
\hspace{3mm}
\begin{minipage}{0.38\linewidth}
    \centering
    \includegraphics[width=1.0\linewidth]{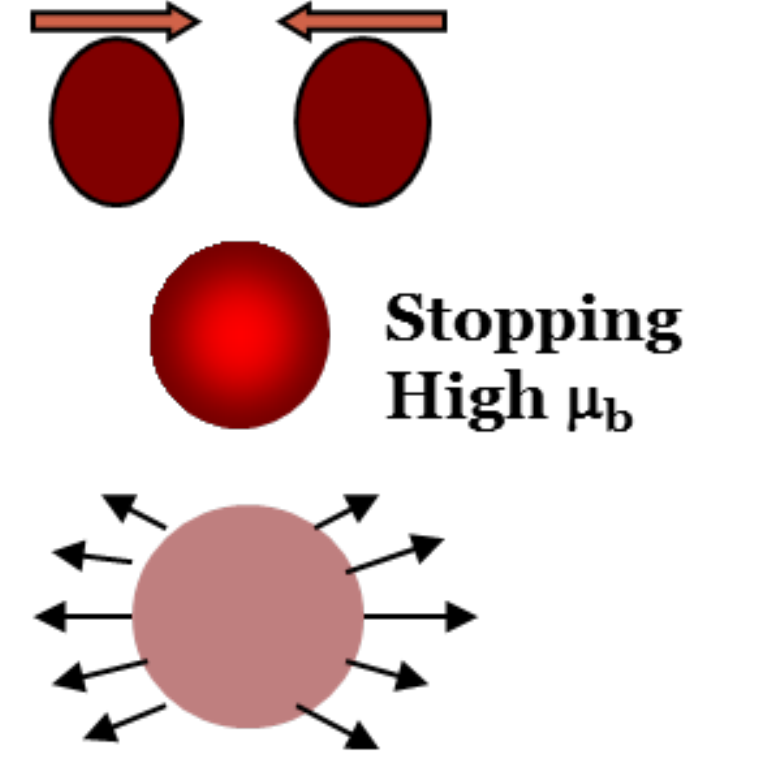}
\end{minipage}
    \caption{\label{fig31} (left) Net-proton $dN/dy$ as a function of rapidity for AGS, SPS and RHIC energies. (right) Schematic drawing of the collision dynamics at AGS (baryon stopping).}
\end{figure}

At the J-PARC energy, which is similar to the AGS energy, one expects
to observe a high density matter created in the midrapidity region.
The left side of Fig.~\ref{fig31} shows the $dN/dy$ of
net-protons~\cite{Bearden:2003hx}.
As going to lower energies from RHIC, SPS to AGS,
the distribution tends to concentrate around $y_{CM}=0$.
At RHIC, the incoming ions are so fast that they pass through
each other, therefore net-protons at the mid-rapidity is low.
At the AGS, the ions are stopped by the friction with other ions
and therefore the net-protons 
becomes large, meaning a high baryon density matter is created.
This phenomena is called as baryon stopping and depicted in the
right side of the Fig.~\ref{fig31}.
From a hadron cascade calculation, JAM~\cite{JAM}, the maximum
baryon density for Au+Au collisions at this energy is estimated
as $\sim$6$\,\rho_0$ and that for U+U collisions as
$\sim$8.6$\,\rho_0$.

\begin{figure}[htbp]
    \centering
    \includegraphics[width=0.9\linewidth]{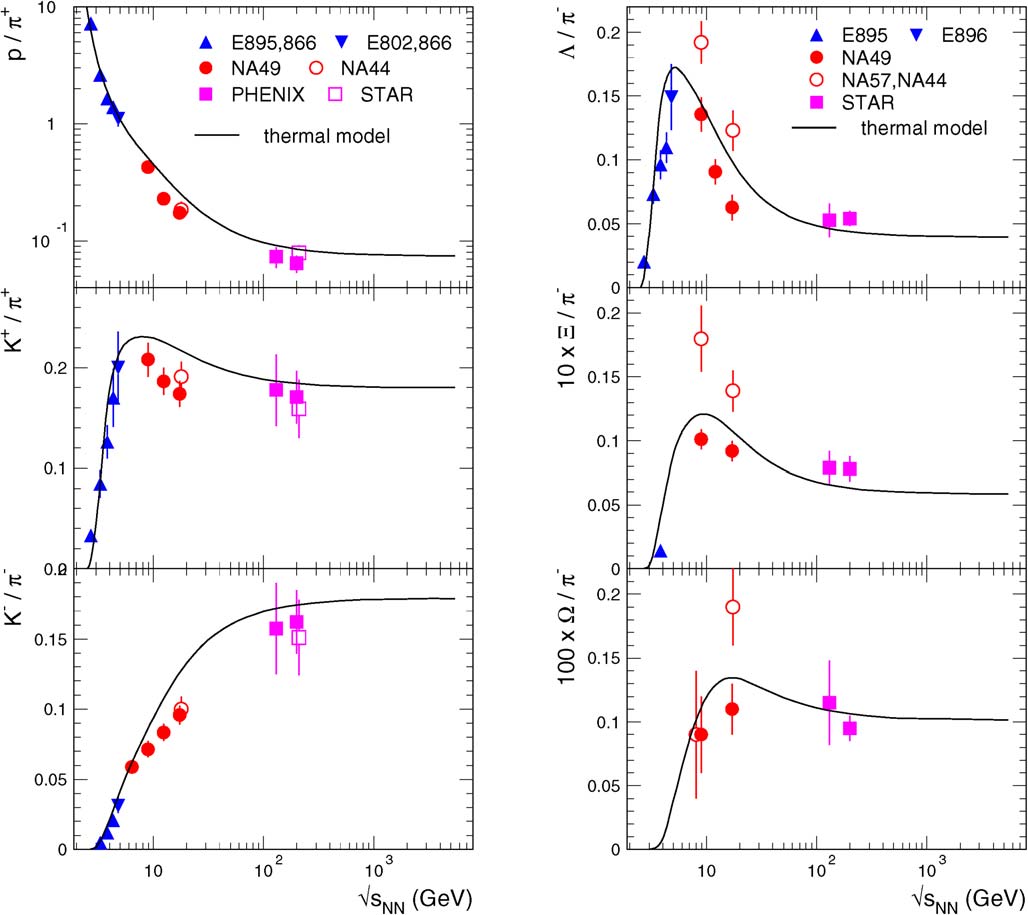}
    \caption{\label{fig32} Compilation of particle ratios as a function
    of collision energies. Enhancement
    of the strangeness particle fraction can be seen in the yield
    increases of $K^+$, $\Lambda$, $\Xi$,
    and $\Omega$ which have $\overline{s}$, with respect to $\pi$ that
    represents the total particle multiplicity.}
\end{figure}

Figure~\ref{fig32} shows the particle ratios as a function of
$\sqrt{s_{NN}}$~\cite{Andronic:2005yp}.
The nominators are numbers of protons or the particles
including open strangeness, while the denominators are the
numbers of $\pi^{+/-}$ which
represent the total particle multiplicities of the events. It is
seen that the ratios whose nominators are strangeness-containing
particles have peaks around $\sqrt{s_{NN}}$=5--8\,GeV. The $K^-/\pi$
ratios are an exception, since the $K^-$ will dominantly be created
from excitation of nucleons. These experimental results show that
the strangeness fraction to the total particles is largest at the
J-PARC energy, suggesting a strangeness-rich matter can be created.

\begin{figure}[htbp]
    \centering
    \includegraphics[width=1.0\linewidth]{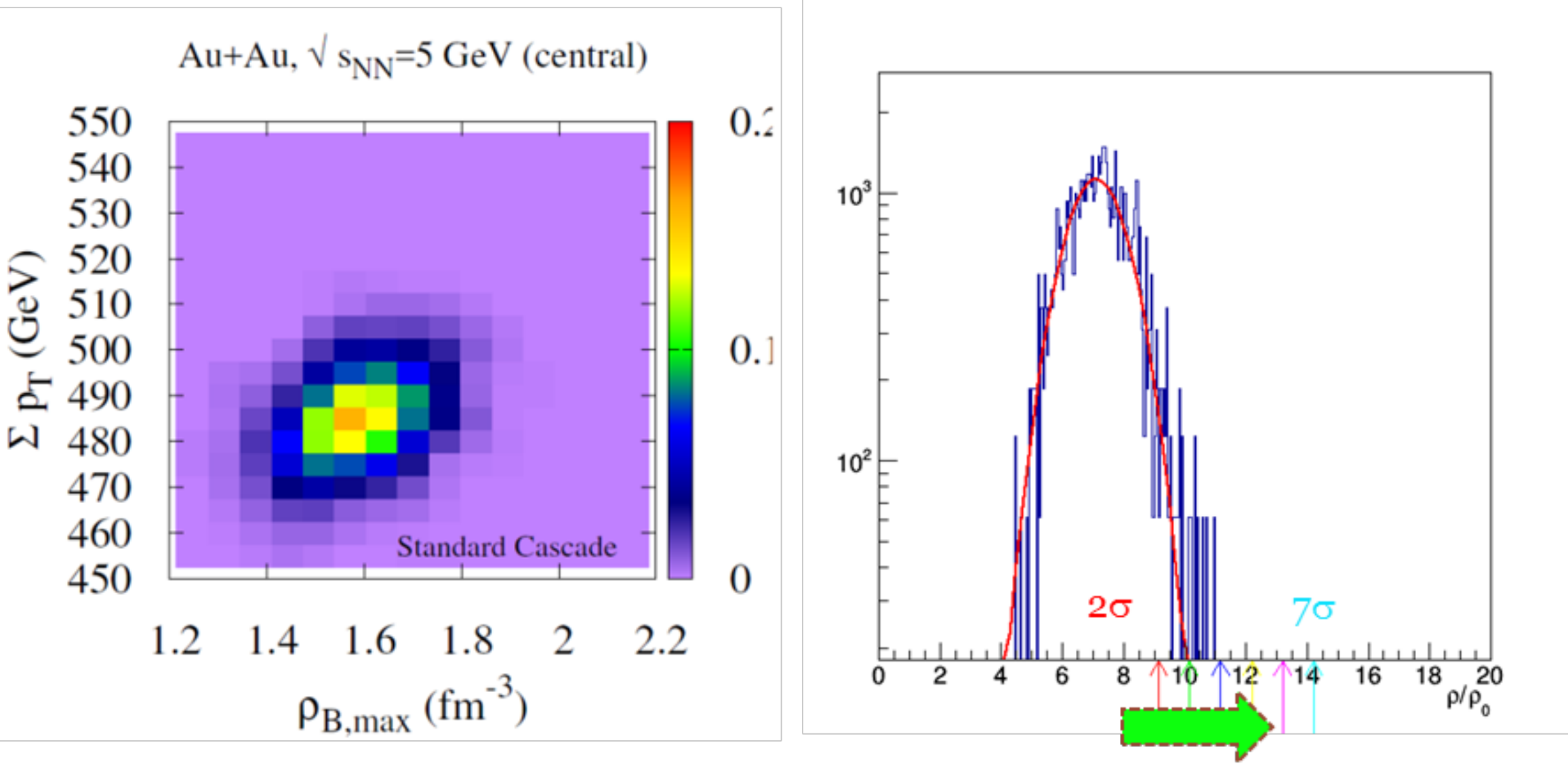}
    \caption{\label{fig33} (left) Correlation between the sum of
    transverse momentum of all the particles and the maximum baryon
    density, $\rho_{B, max}$, for given events, obtained by a standard
    hadron cascade simulation, JAM. (right) $\rho/\rho_0$ distribution
    from the same simulation. One order higher statistics will push the
    reach of $\rho/\rho_0$ higher by 1\,$\sigma$.}
\end{figure}

In order to observe a new phenomena at this energy, one has to
invent a new event selection scheme. Traditionally, the heavy
ion physicist chose most violent and less by using the variables
called centrality. The centrality can be associated with impact
parameters using a Glauber Monte Carlo calculation. However, it
doesn't guarantee the selection of very high baryon density events.
Here we demonstrate a new event selection from the study
using the JAM code. Figure~\ref{fig33} show the sum of the
transverse momentum of all the particles produced in an event
plotted against the maximum baryon density in the event.
We can see a correlation between two quantities. The right side of the
Fig~\ref{fig33} shows the distribution of $\rho/\rho_0$,
where the $\rho_0$ is the normal nuclear density. It is
shown that a one order higher statistics can push the
reach of $\rho/\rho_0$ higher by 1\,$\sigma$.

\begin{figure}[htbp]
\begin{minipage}{0.48\linewidth}
    \centering
    \vspace*{-5mm}
    \includegraphics[width=1.0\linewidth]{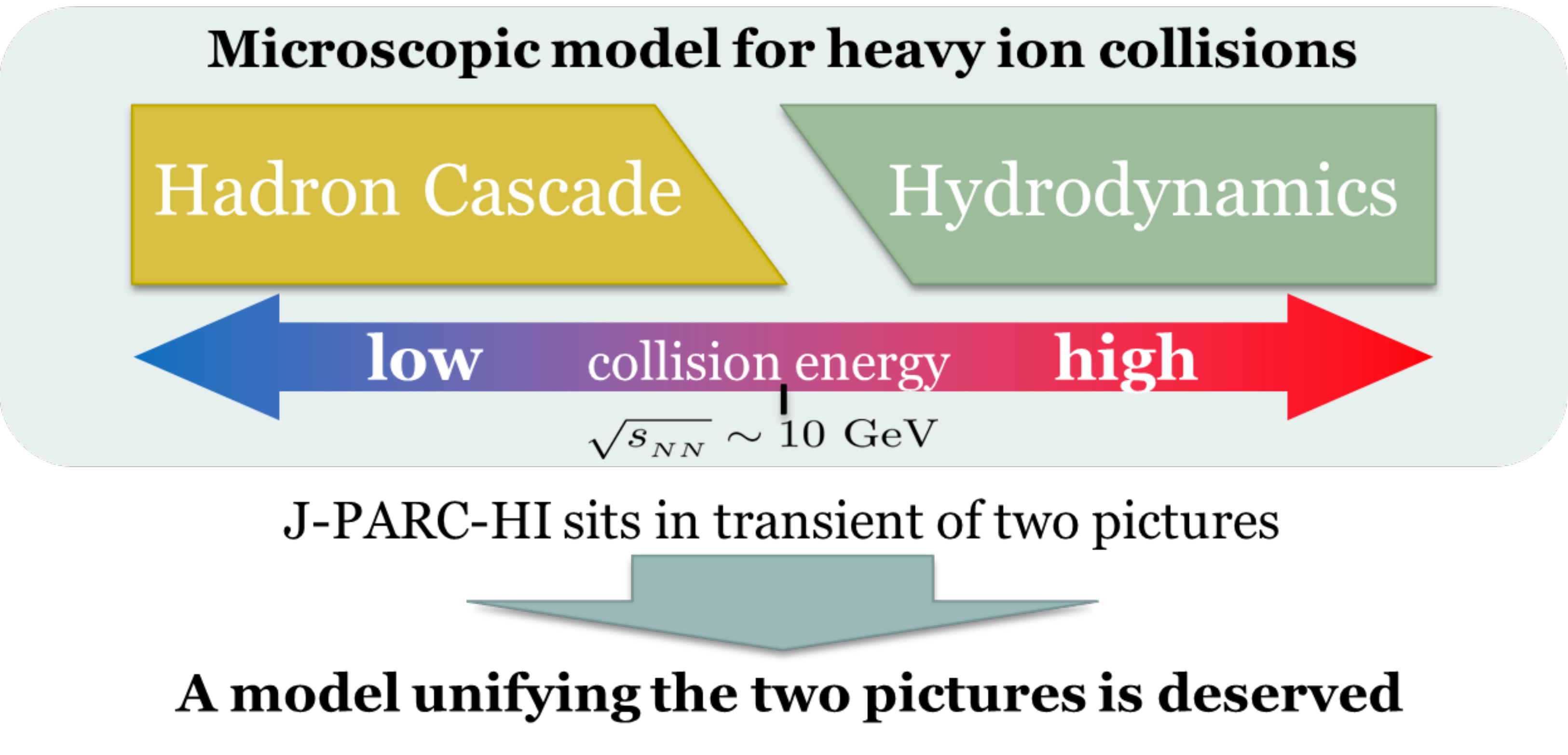}\\
    \vspace*{8mm}
    \includegraphics[width=0.7\linewidth]{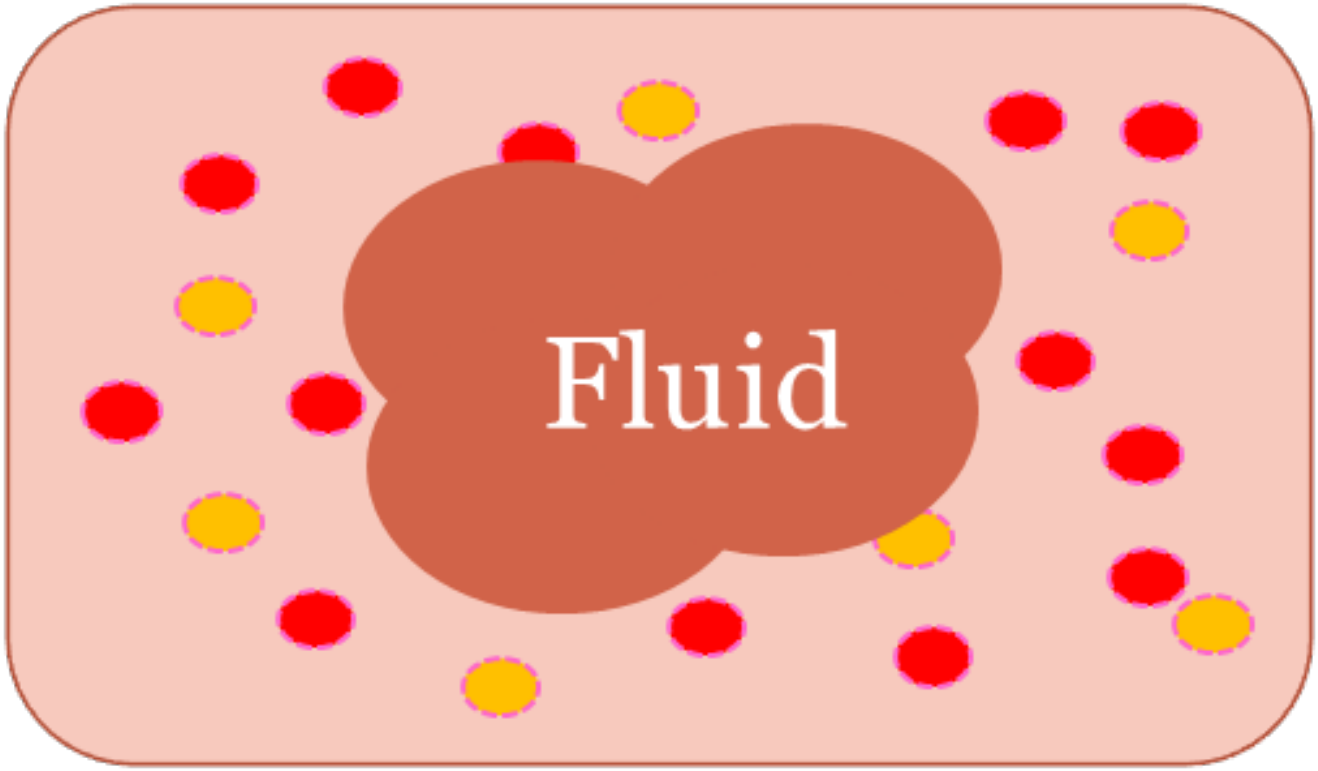}
\end{minipage}
\hspace{3mm}
\begin{minipage}{0.48\linewidth}
    \centering
    \includegraphics[width=1.0\linewidth]{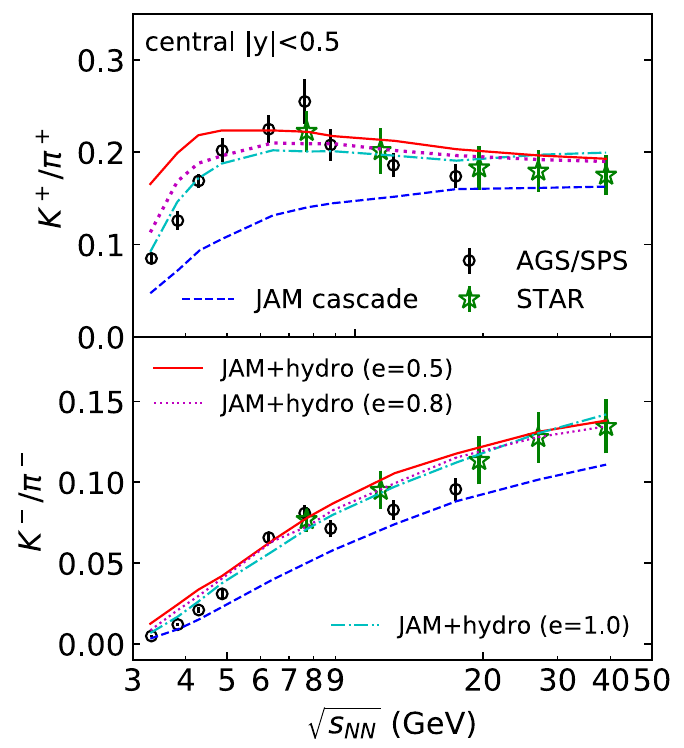}
\end{minipage}
\caption{\label{fig35} A theoretical effort to understand the collision
dynamics at the J-PARC-HI energy. The dynamics should be described in
a mixed framework of hadron cascade and hydrodynamics. The right plots
show the particle ratios from the conventional hadron cascade framework
and the newly proposed mixed framework.}
\end{figure}

This new experimental project stimulated the theory community as well.
The matter created in high energy collisions can be described by
a hydrodynamical model rather well, while the one in low energy
collisions can mainly described by a hadron cascade model as
depicted in the left side of the Figure~\ref{fig35}.
The J-PARC-HI energy sits in transient of two pictures, therefore
a model unifying the two pictures is desired. Recently, the
Japanese theoretical community that has been working on heavy
ion physics and is interested in the J-PARC-HI project
invented a new such model~\cite{Akamatsu:2018olk}. The model
introduced a threshold parameter $e$ which
corresponds to the fluid energy density. For the region whose
density is above $e$, the hadrons are turned into a parton
fluid and plugged into a hydrodynamical framework. The region
where the density is below $e$, the parton fluid is 
"cooled down" to hadrons. The right side of the
Fig~\ref{fig35} shows the $K^+/\pi^+$ and $K^-/\pi^-$
as a function of $\sqrt{s_{NN}}$ compared with the calculations
from the conventional JAM hadron cascade code as well as the
newly proposed unified framework with two $e$ values.
It is clearly seen that the unified models give better
description of the data~\cite{Akamatsu:2018olk}.

\subsection{Observables: Dileptons}
\begin{figure}[htbp]
    \centering
    \includegraphics[width=0.6\linewidth]{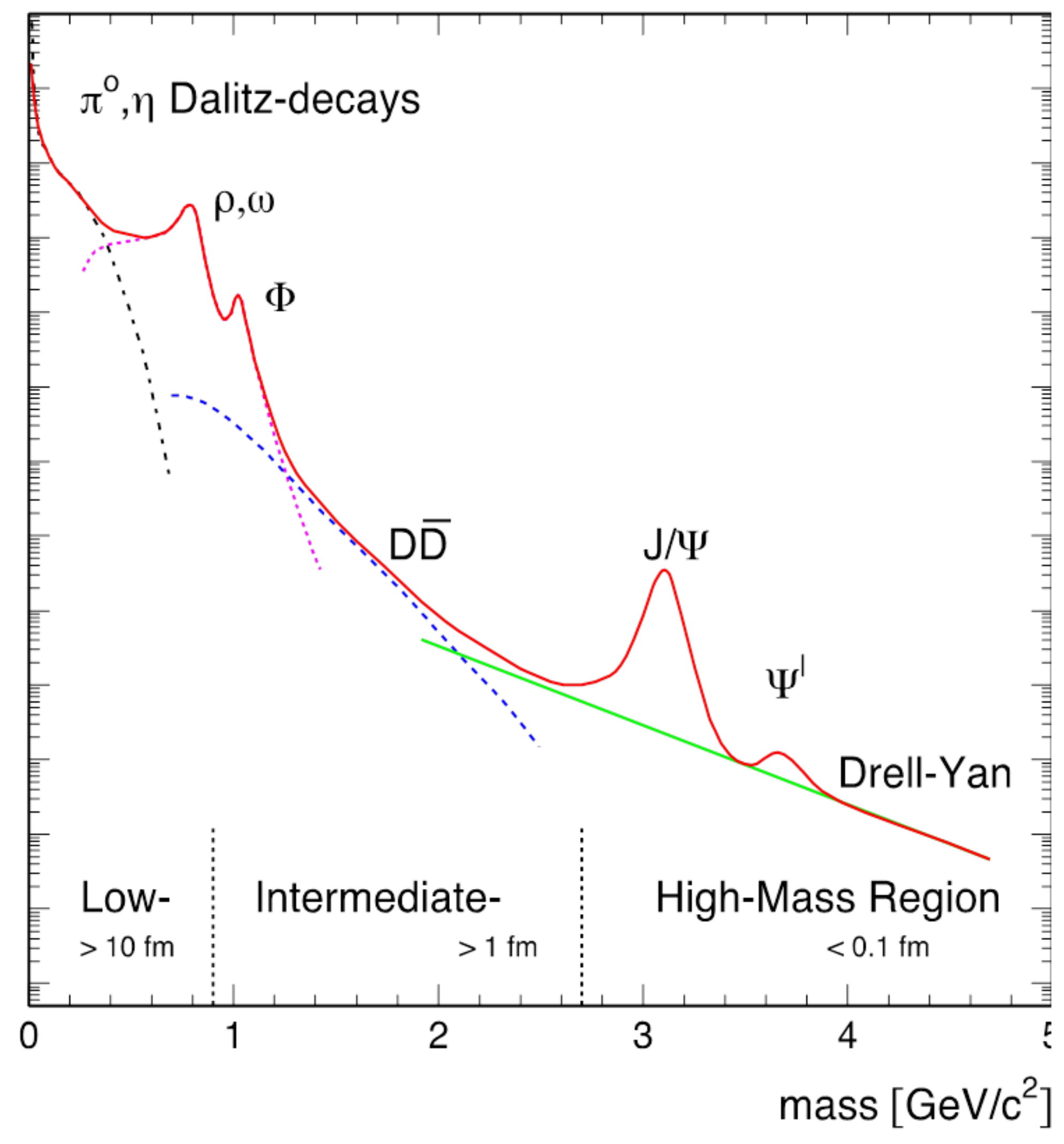}
    \caption{\label{fig36} Schematic drawing of the lepton-pair
    (electron-positron pair in this case) mass distribution from
    various contributions.}
\end{figure}

Dileptons are one of the important observables that have not been
measured at the AGS/J-PARC-HI energies. Figure~\ref{fig36} shows
the schematic plot showing the various contribution decaying into
lepton-pairs~\cite{Rapp:1999ej}.
The hadrons contributing to the low mass region
(0.3$<m_{ll}<$1\,GeV: $\rho$, $\omega$, and $\phi$)
are of course of interest for long time as they are promising
observables to probe chiral symmetry restoration effect.
A recent study claims that the moments of the dilepton mass spectrum:
\[\int dm_{ll} N(m_{ll})m_{ll}^n\ \ (n=1,2,...)\]
can be related to the quark/gluon condensate of the system~\cite{Hayano:2008vn}.
Since the uncertainty will become larger as going to a lower mass region
due to a large background, a high statistics experiment such as
J-PARC-HI is suitable to tackle this measurement. The mass
region lower than $m_{ll}<$100\,MeV is told to be the window
associated with the soft mode that arises in the transition
to the color-superconductivity phase~\cite{Kitazawa:2001ft}.
The one above $\phi$ is dominated by leptons decaying from
$c\overline{c}$ at the energy of SPS and above, where the charm
production is copious. However, the $c\overline{c}$ production
is highly suppressed at J-PARC energy because of kinematical
limit. Therefore, this mass region
is ideal for observing the thermal radiation from the matter at
this energy~\cite{Song:2018xca}.

\subsection{Observables: fluctuation}
\begin{figure}[htbp]
    \centering
    \includegraphics[width=0.9\linewidth]{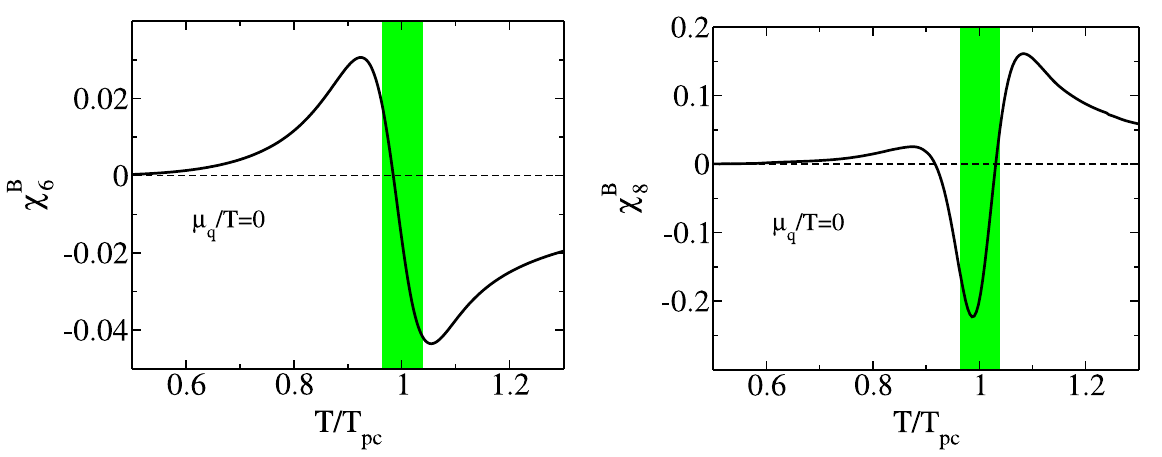}
    \caption{\label{fig37} Sixth (left) and eighth (right) moments of the net-baryon fluctuation around the phase transition temperature ($T_{pc}$). It is clearly seen that the moments vary significantly around $T_{pc}$.}
\end{figure}

The fluctuation of conserved quantities has been of interest since
it is sensitive to the transition of the phases, especially,
the critical point. One of the typical observables
is the net-baryon fluctuation. In reality, one can only measure
the net-protons due to difficulty of measuring neutrons. It is,
however, one of the goals at J-PARC-HI to measure the neutrons and
thus the genuine net-baryon fluctuations. Another interesting
observable is the higher order cumulants which are more sensitive
to the phase transition. Both RHIC and LHC experiments have
measured up to fourth order moments (kurtosis), but the sixth and
eighth moments are said to even more sensitive to the chiral phase
transition as shown in Fig.~\ref{fig37}~\cite{Friman:2011pf}.
It is shown that the moments vary significantly around $T_{pc}$.
It should be noted that around one order magnitude higher statistics
is necessary to observe moments of one order higher. Therefore,
a very high statistics experiment such as J-PARC-HI is ideal to
measure such an observable.

\subsection{Observables: flow}
\begin{figure}[htbp]
\begin{minipage}{0.30\linewidth}
    \centering
    \includegraphics[width=1.0\linewidth]{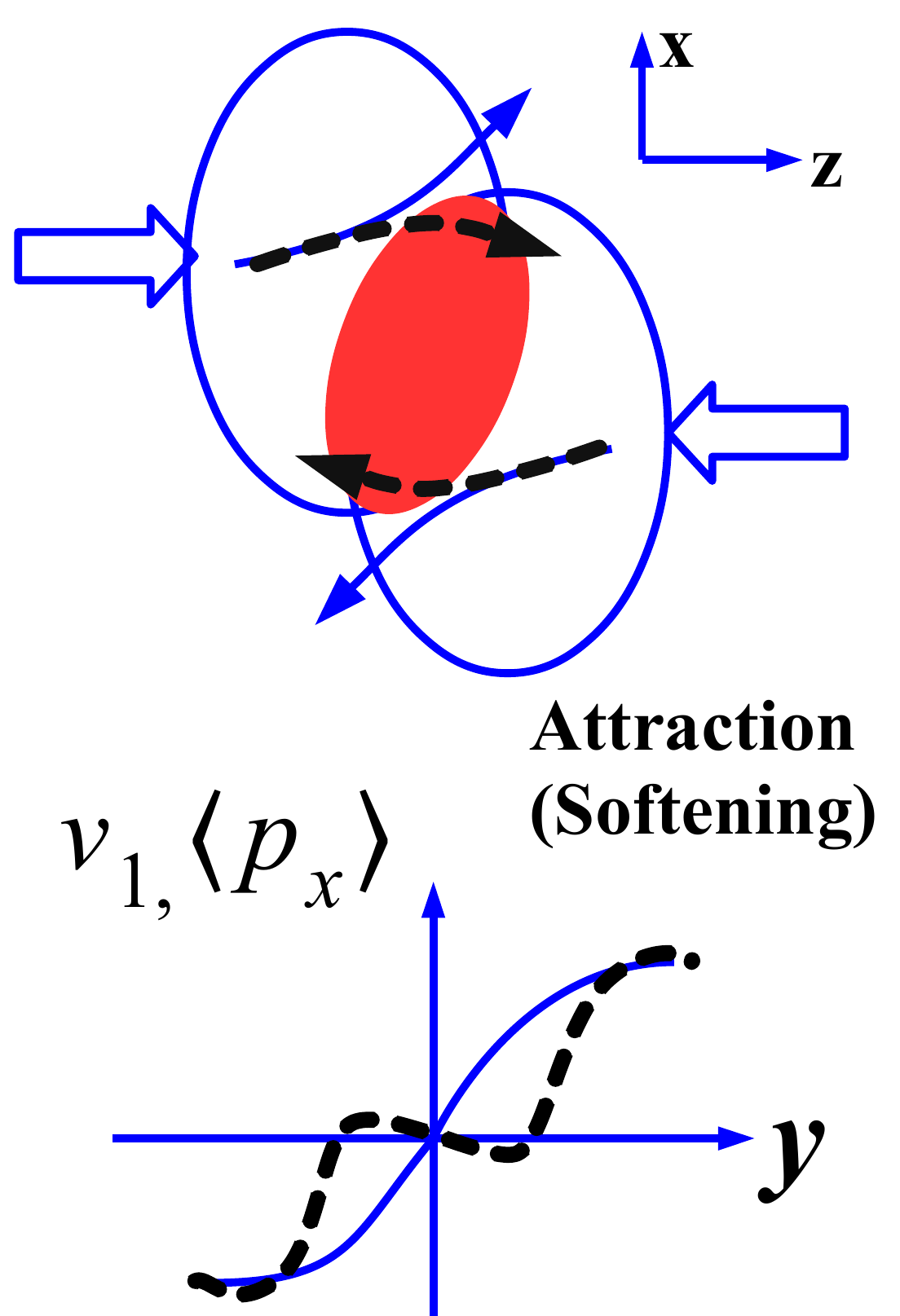}
\end{minipage}
\hspace{4mm}
\begin{minipage}{0.65\linewidth}
    \centering
    \includegraphics[width=1.0\linewidth]{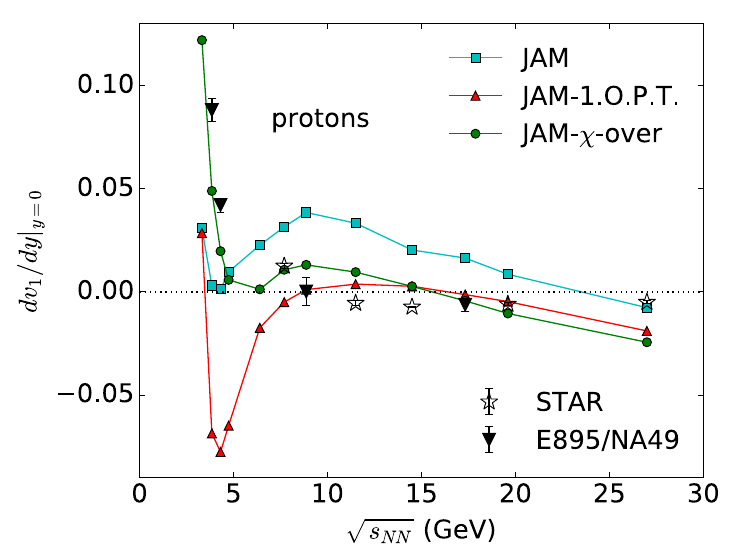}
\end{minipage}
    \caption{\label{fig38} (left) Schematic drawing of the possible effect of EOS softening to the directed flow ($v_1$, $dv_1/dy$). (right) Directed flow slopes as a function of $\sqrt{s_{NN}}$ together with hadron
    cascade calculations, JAM, with no phase transition,
    first order transition, and cross-over transition.}
\end{figure}

The particle flow is sensitive to the equation of state (EOS) of
the matter created in the early stages of the collisions.
Depending on the order of
the phase transition, the softening process will become different
which results in a different flow strength. One of the flow
components that are said to be sensitive is the directed flow
($v_1$) and its slope as a function of rapidity ($dv_1/dy$)
around the mid-rapidity ($y=0$). When softening of the EOS
occurs, the nucleons in the participant region are attracted
(shown in the left side of Fig.~\ref{fig38}) and therefore
$dv_1/dy$ becomes negative for nucleons. 
This is demonstrated by calculations of $dv_1/dy$ for protons
using JAM for the case of no phase transition, first phase phase
transition, and cross-over phase transition as shown
in the right side of Fig.~\ref{fig38}~\cite{Nara:2016hbg}.
It is clearly seen that
in the cases of no phase transition or cross-over phase
transition, the $dv_1/dy$ stays positive over the energies,
while in the case of first order phase
transition, the sign changes at the energy where the
transition is expected to occur.
In addition to the directed flow, high order flows
are interesting in the similar sense of measuring the
higher moments in the fluctuation measurement of the
conserved quantities. It is also
true that going to an order higher flow requires an
order higher statistics where J-PARC-HI can contribute
significantly.

\subsection{Observables: hypernuclei and others}
The exploration of the multi-strange hypernuclei ($S\geq3$)
will become
possible at the beam rapidity; the hypernuclei will be
looked for in the beam fragment. The lifetime and the
magnetic moment are of interest of the hypernuclei
physics community since they provide information of the
structure of the hypernuclei~\cite{Lambda-magmom,Sigma-magmom}.
The reaction cross-section should also be measured to
obtain an idea on the size of the hypernuclei.

The hadron-hadron correlations can serve to many interesting
topics of the hadron physics. For instance, the
$\Lambda$-$\Lambda$ correlation has been looked at 
from the point of view of dibaryon search~\cite{Adamczyk:2014vca},
since the bound state can be detected by the strength of this
correlation. The correlation can also be used to explore changes
of the baryon-baryon interaction in the high baryon density
environment. A recent Lattice calculation found that the
$\Omega$-$\Omega$ creates a bound state~\cite{Gongyo:2017fjb}.
Therefore, this correlation can be utilized in the same sense.

More exotics can be searched with very high statistics at J-PARC.
Both Kaonic nucleus and strangelets
have been of interest of hadron physics, since they are new
forms of the hadrons whose natures are barely known. Exploring
multi-quark states such as 6, 8, and/or 10 quark states will
shed light on the nature of the strong interaction as well as
the transition to a mini-QGP state.

\section{Proposed detectors}
The detector has to be capable of
handling hits at very high rate, especially for the particle tracking
part. At same time, a large acceptance will be needed for high
statistics inclusive measurement of the event-by-event fluctuation.
Our plan is to build and reconfigure detectors in three stages,
preceded by a day-1 precursor experiment that will be described later.
We will employ a high speed continuous data taking system combined
with an online and a semi-online trigger system at all stages,
in order to reduce the
amount of data to be recorded on disk and thus to reduce the time to
extract and analyze interesting events.

\begin{figure}[htbp]
%\begin{minipage}{0.54\linewidth}
    \centering
    \includegraphics[width=0.9\linewidth]{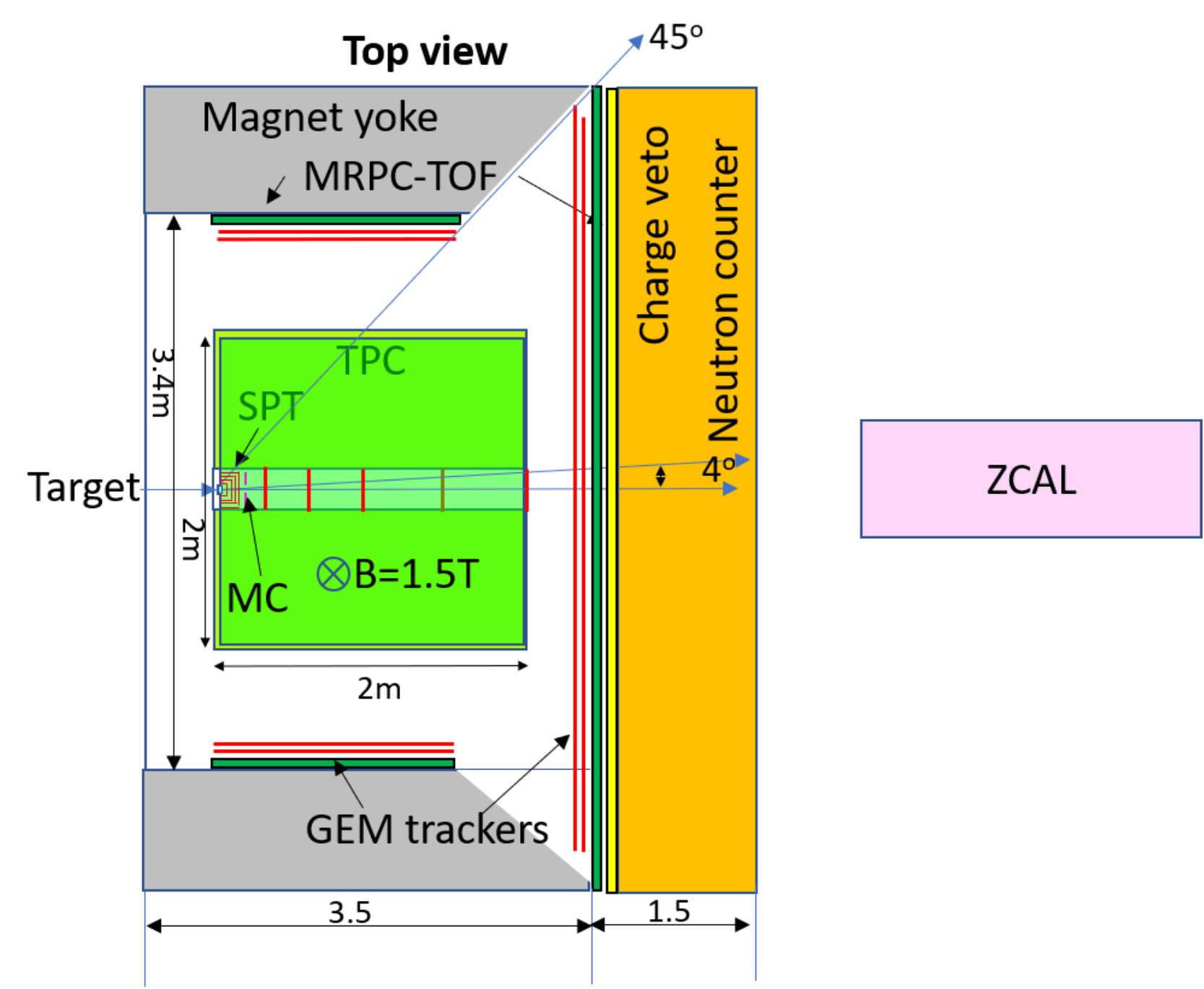}
%\end{minipage}
\caption{\label{fig40} First stage experiment setup focusing on the haronic observables (hadron spectra, flow, and fluctuation).}
\end{figure}

The first stage focuses on the observables that can be measured
at lower intensity ($10^6$\,Hz).
This is basically the soft particle production, and its
correlation/fluctuation. The Fig.~\ref{fig40} shows
a dipole hadron spectrometer to be operated at the first stage of
the experiment.
%\hspace{3mm}
%{\color{red}\vline\vline\vline\vline}
\begin{figure}[htbp]
%\begin{minipage}{0.40\linewidth}
    \centering
    \includegraphics[width=0.6\linewidth]{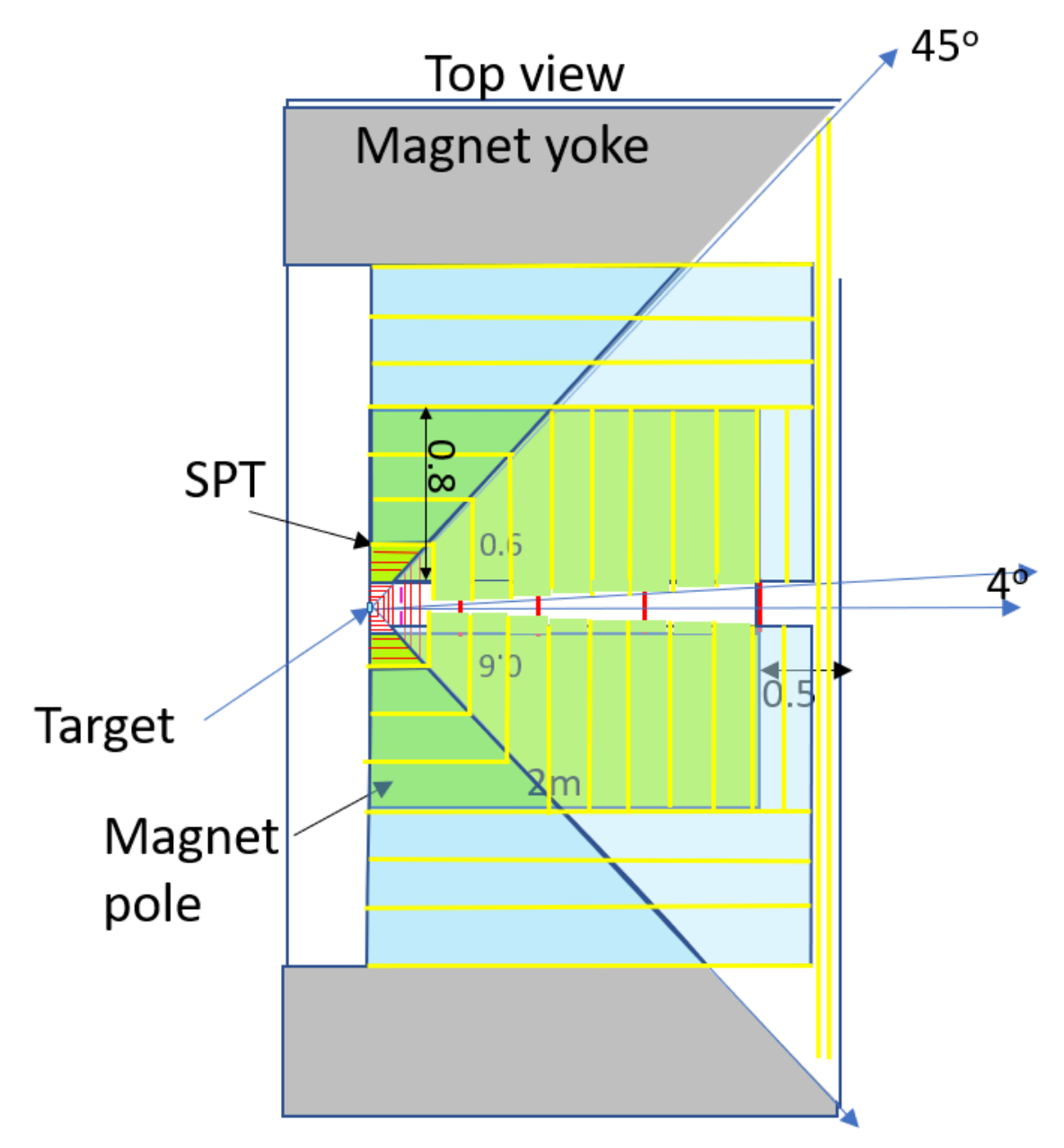}
%\end{minipage}
\caption{\label{fig41} Second stage experimental setup focusing on dilepton (dimuon) measurement.}
\end{figure}
The main tracking device consists of silicon pixel detectors
($\theta$$<$40$^\circ$) and a time projection chamber
(TPC, $\theta$$>$40$^\circ$)
which have an exclusive zenith coverage each but have a full-azimuth
coverage. The tracking device is then followed by a time-of-flight
(TOF) detector with multi-resistive proportional chamber (MRPC) and a
neutron counter. The centrality will be determined by multiplicity
counters located in front of the silicon pixel detectors and a
zero-degree hadronic calorimeter located in the downstream of the beam.

\begin{figure}[htbp]
    \centering
    \includegraphics[width=0.9\linewidth]{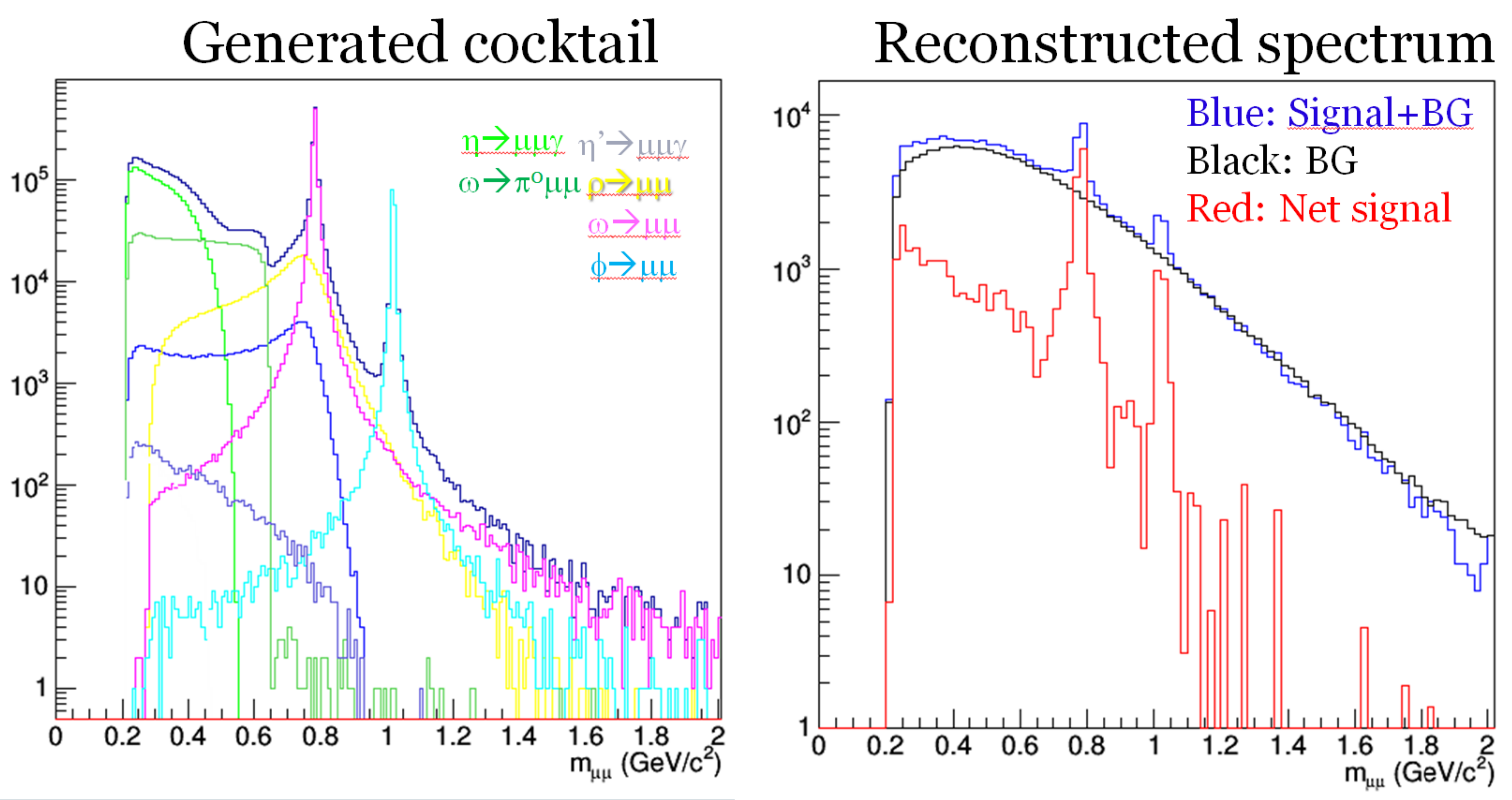}
    \caption{\label{fig42} Simulated dimuon mass spectra from various contributions processed through a GEANT detector simulation at J-PARC-HI. (left) Generated dimuon spectra by the modified EXODUS event generator. (right) Reconstructed dimuon spectra from the dimuon signals embedded into JAM events. The branching ratios of hadrons decaying into $\mu^+\mu^-$ are enhanced by a factor of 1000.}
\end{figure}

The second stage experiment will perform
dilepton measurement via $\mu\mu$ channel with a spectrometer
shown in the Fig.~\ref{fig41}. The TPC will be replaced by a dimuon
tracker consisting of GEM trackers sandwiched with Pb absorbers.
In order to improve the momentum and track resolution, 7-layers
of forward and barrel silicon pixel detectors will also be installed.
We will be able to operate the detector at the rate up to $10^7$\,Hz.
We have generated simulated event of hadrons decaying into muons
by a modified EXODUS event generator~\cite{Adare:2009qk}
originally developed by the
PHENIX experiment as shown in the left side of the Fig.~\ref{fig42}.
We then embedded the dimuons into the JAM events, processed them through a
GEANT4 detector simulation software~\cite{Agostinelli:2002hh}. 
The reconstructed dimuon
mass spectra which is shown in the right side of the Fig.~\ref{fig42}.
The $\pi$ and $K$ hadrons generated by the JAM were decayed into
muons by the GEANT4.
The collision energy is set to be $\sqrt{s_{NN}}$=4.5\,GeV. The
particles passing through the 4$\lambda_I$ absorbers were
identified as muons. 
The peaks of $\rho/\omega$ and $\phi$ are visible in the continuous
background. Note that in this simulation, the branching ratios of
hadrons decaying into $\mu^+\mu^-$ are enhanced by a factor of 1000
in order to earn enough statistics for the signal dimuons.

\begin{figure}[htbp]
    \centering
    \includegraphics[width=1.0\linewidth]{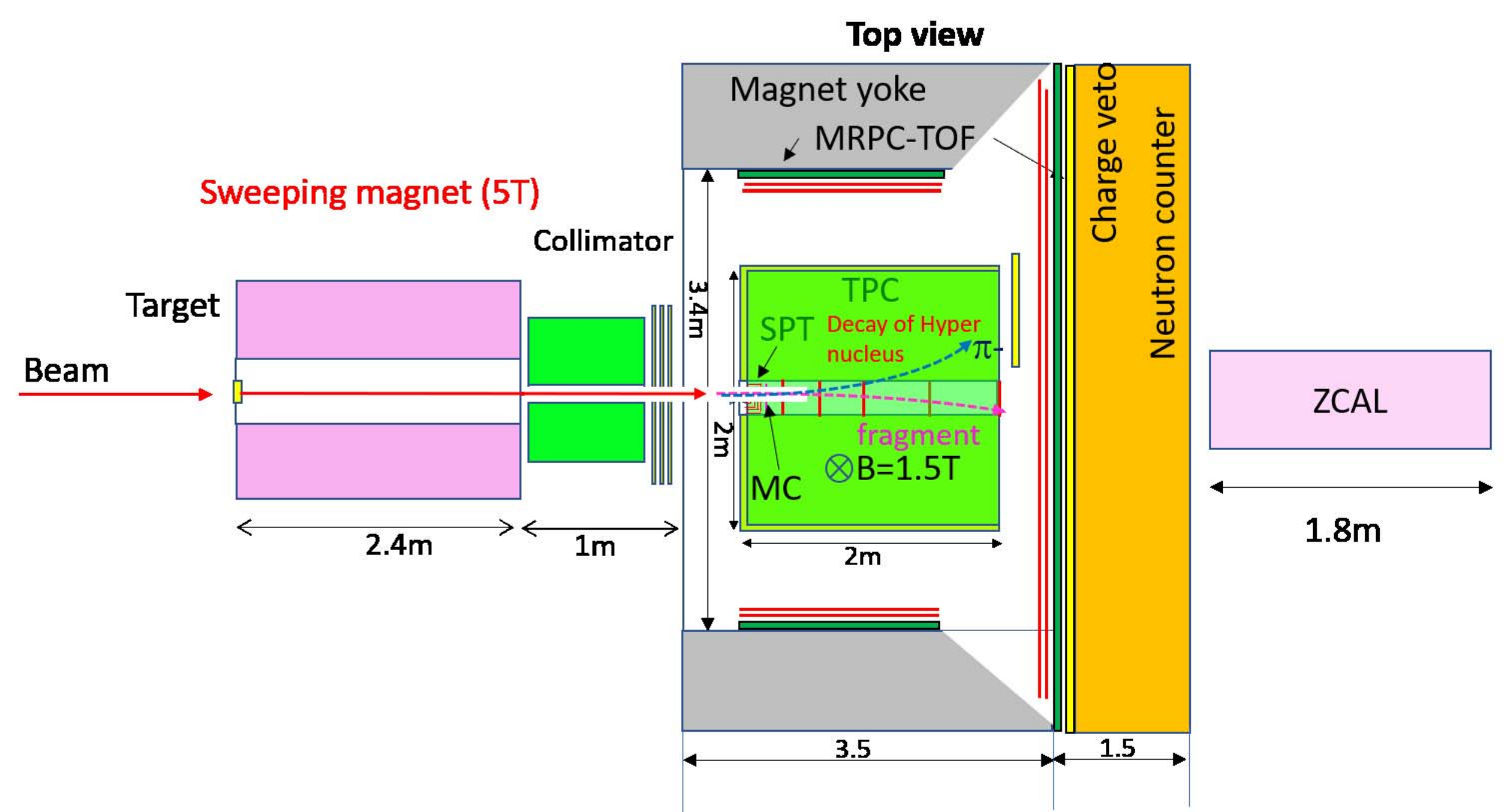}\\
    \vspace*{5mm}
    \includegraphics[width=1.0\linewidth]{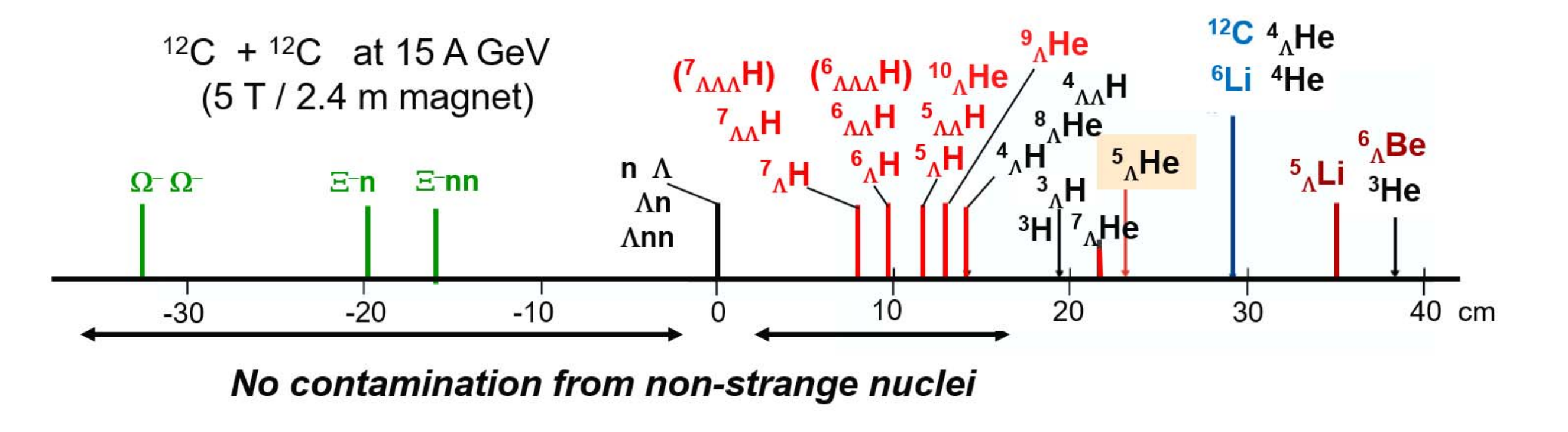}
    \caption{\label{fig43} Experimental setup focusing on the hypernuclei (top).
    The projected position will be different depending on the masses of
    the hypernuclei (bottom). The ones arriving at the negative position
    with respect to the beam position will be negatively charged nuclei ever measured.}
\end{figure}

The third stage experiment focuses on hypernulei 
measurement. The setup is shown in Fig.~\ref{fig43}.
The beam will be bombarded to the target followed by the
sweeping magnet of around 5\,T and the collimator. The beam
particles will be swept away by the magnet and the collimator,
and therefore only beam fragments which include hypernuclei
will be introduced to the tracking part.
The $\pi$'s coming from the weak decay of the hypernuclei will
be tracked and identified by the TPC and the MRPC-TOF, and
will be used as tags for hypernuclei events.
The projected arrival positions of the hypernuclei at the exit
of the collimator will be reconstructed and used for identifying
the species of the hypernuclei. The correlation of the projected
positions and the species of the hypernuclei
is shown in the bottom panel of Fig.~\ref{fig43}. This
experiment will require very high rate ($\sim10^8$\,Hz)
since the production cross-section of the hypernuclei is
significantly small.

\section{Project status}
\begin{figure}[htbp]
    \centering
    \includegraphics[width=1.0\linewidth]{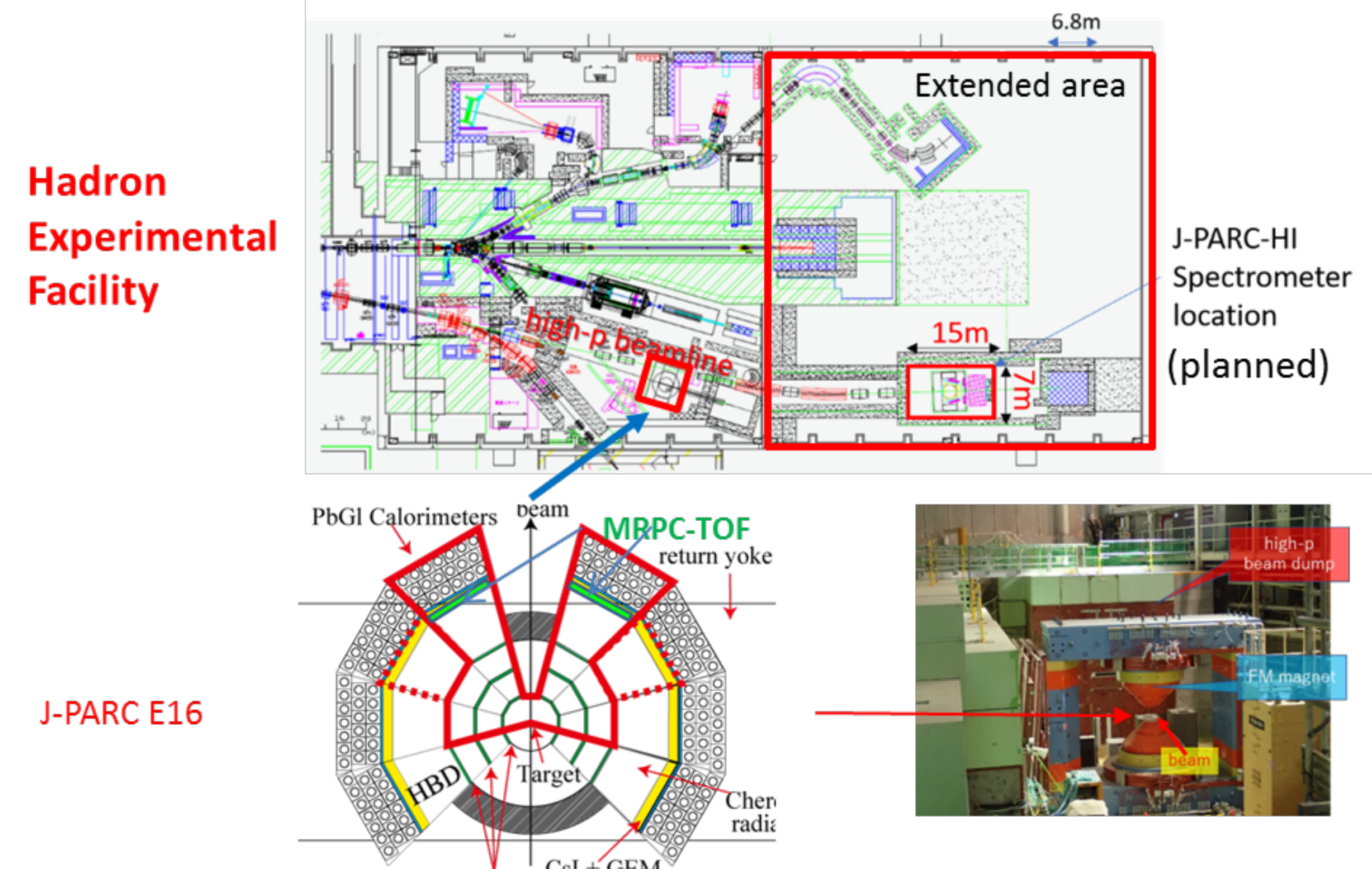}
    \caption{\label{fig44} Candidate location for the J-PARC-HI experimental setup. The E16 experiment sitting upstream of the J-PARC-HI beam line is an approved experiment for electron pair measurement, and may be utilized as the day-1 precursor experiment for heavy ion collisions.}
\end{figure}

The plan of upgrading J-PARC to J-PARC-HI for accelerating
heavy ions has been intensively discussed
in the Japan Atomic Energy Agency (JAEA) and the J-PARC.
The upgrade includes building an ion source,
a linac, and a booster to generate and accelerate heavy
ions for injecting to the RCS. The cost for the accelerator
part has been estimated
to be $\sim\$$150\,M. We have submitted this proposal
for the master plan of science council of Japan, which is
the 10-year plan staring from 2020. If successful, the project
will be admitted as a prioritized plan and become subject
to budget discussion.
As for the experimental side, we have identified the
location for these detector systems in the J-PARC
hadron experimental hall as shown in Fig.\ref{fig44}.
In the current experimental hall, the primary proton beam coming from the left
side of the picture is bombarded to a target and only the secondary
particles are transported to experiments. A new beam line called 
high momentum beam line is being built in the hall to
transport the
primary beam to experiments. The experimental area
is currently limited to the left half of the drawing, but
there is a plan to extend the area as shown in the right
half of the drawing. The J-PARC-HI will be built
on the high momentum beam line in the extended area.
On this beam line, there is an approved experiment,
E16~\cite{E16}, being built for lepton-pair
measurement using proton-nucleus collisions. We are
considering to add time-of-flight and centrality
detectors to the E16 detector and perform the day-1
precursor experiment with heavy ion beams until the
extended area becomes available.

\section{Summary and near term prospect}
In this paper, we discussed the physics perspective of
the high density matter study.
We introduced the J-PARC-HI project which is a unique lab
to tackle the high density matter physics. The world highest
rate of heavy ion beam of 10$^{11}$\,Hz is aimed at J-PARC-HI
which makes us possible to perform measurements of hadrons, 
fluctuation of conserved quantities, dileptons,
multi-strange hypernuclei. New event selections were also
discussed to reach a highest baryon density on the ground.
A white paper and a letter of intent were written and
submitted in 2016~\cite{jparcweb:2016}, and the proposal
for the master plan of Science Council of Japan
was submitted in Mar 2019. Taking the budget situation into
account, we anticipate that the
day-1 precursor experiment may be performed in 2026 at earliest.
The staged detector systems will be built in the
extended area of the hadron experimental hall when
it becomes available.

\bibliographystyle{JHEP}
\bibliography{SakaguchiCPOD18}

\providecommand{\href}[2]{#2}\begingroup\raggedright\begin{thebibliography}{10}

\bibitem{fair}
{\emph{FAIR baseline technical report} {\bfseries 3a} (2006) }.

\bibitem{nica}
{\emph{Design and construction of nuclotron-based ion collider facility (NICA)
  conceptual design report} (2008) }.

\bibitem{Andronic:2010qu}
A.~Andronic, P.~Braun-Munzinger, J.~Stachel and H.~Stocker, \emph{{Production
  of light nuclei, hypernuclei and their antiparticles in relativistic nuclear
  collisions}},
  \href{https://doi.org/10.1016/j.physletb.2011.01.053}{\emph{Phys.Lett.}
  {\bfseries B697} (2011) 203}
  [\href{https://arxiv.org/abs/1010.2995}{{\ttfamily 1010.2995}}].

\bibitem{BraunMunzinger:1994iq}
P.~Braun-Munzinger and J.~Stachel, \emph{{Production of strange clusters and
  strange matter in nucleus-nucleus collisions at the AGS}},
  \href{https://doi.org/10.1088/0954-3899/21/3/002}{\emph{J. Phys.} {\bfseries
  G21} (1995) L17} [\href{https://arxiv.org/abs/nucl-th/9412035}{{\ttfamily
  nucl-th/9412035}}].

\bibitem{Bearden:2003hx}
{\scshape BRAHMS} collaboration, \emph{{Nuclear stopping in Au + Au collisions
  at $\sqrt{s_{NN}}$=200\,GeV}},
  \href{https://doi.org/10.1103/PhysRevLett.93.102301}{\emph{Phys. Rev. Lett.}
  {\bfseries 93} (2004) 102301}
  [\href{https://arxiv.org/abs/nucl-ex/0312023}{{\ttfamily nucl-ex/0312023}}].

\bibitem{JAM}
Y.~Nara, N.~Otuka, A.~Ohnishi, K.~Niita and S.~Chiba, \emph{{Study of
  relativistic nuclear collisions at AGS energies from p + Be to Au + Au with
  hadronic cascade model}},
  \href{https://doi.org/10.1103/PhysRevC.61.024901}{\emph{Phys.Rev.} {\bfseries
  C61} (2000) 024901} [\href{https://arxiv.org/abs/nucl-th/9904059}{{\ttfamily
  nucl-th/9904059}}].

\bibitem{Andronic:2005yp}
A.~Andronic, P.~Braun-Munzinger and J.~Stachel, \emph{{Hadron production in
  central nucleus-nucleus collisions at chemical freeze-out}},
  \href{https://doi.org/10.1016/j.nuclphysa.2006.03.012}{\emph{Nucl. Phys.}
  {\bfseries A772} (2006) 167}
  [\href{https://arxiv.org/abs/nucl-th/0511071}{{\ttfamily nucl-th/0511071}}].

\bibitem{Akamatsu:2018olk}
Y.~Akamatsu, M.~Asakawa, T.~Hirano, M.~Kitazawa, K.~Morita, K.~Murase et~al.,
  \emph{{Dynamically integrated transport approach for heavy-ion collisions at
  high baryon density}},
  \href{https://doi.org/10.1103/PhysRevC.98.024909}{\emph{Phys. Rev.}
  {\bfseries C98} (2018) 024909}
  [\href{https://arxiv.org/abs/1805.09024}{{\ttfamily 1805.09024}}].

\bibitem{Rapp:1999ej}
R.~Rapp and J.~Wambach, \emph{{Chiral symmetry restoration and dileptons in
  relativistic heavy ion collisions}},
  \href{https://doi.org/10.1007/0-306-47101-9_1}{\emph{Adv. Nucl. Phys.}
  {\bfseries 25} (2000) 1}
  [\href{https://arxiv.org/abs/hep-ph/9909229}{{\ttfamily hep-ph/9909229}}].

\bibitem{Hayano:2008vn}
R.~S. Hayano and T.~Hatsuda, \emph{{Hadron properties in the nuclear medium}},
  \href{https://doi.org/10.1103/RevModPhys.82.2949}{\emph{Rev. Mod. Phys.}
  {\bfseries 82} (2010) 2949}
  [\href{https://arxiv.org/abs/0812.1702}{{\ttfamily 0812.1702}}].

\bibitem{Kitazawa:2001ft}
M.~Kitazawa, T.~Koide, T.~Kunihiro and Y.~Nemoto, \emph{{Precursor of color
  superconductivity in hot quark matter}},
  \href{https://doi.org/10.1103/PhysRevD.65.091504}{\emph{Phys. Rev.}
  {\bfseries D65} (2002) 091504}
  [\href{https://arxiv.org/abs/nucl-th/0111022}{{\ttfamily nucl-th/0111022}}].

\bibitem{Song:2018xca}
T.~Song, W.~Cassing, P.~Moreau and E.~Bratkovskaya, \emph{{Open charm and
  dileptons from relativistic heavy-ion collisions}},
  \href{https://doi.org/10.1103/PhysRevC.97.064907}{\emph{Phys. Rev.}
  {\bfseries C97} (2018) 064907}
  [\href{https://arxiv.org/abs/1803.02698}{{\ttfamily 1803.02698}}].

\bibitem{Friman:2011pf}
B.~Friman, F.~Karsch, K.~Redlich and V.~Skokov, \emph{{Fluctuations as probe of
  the QCD phase transition and freeze-out in heavy ion collisions at LHC and
  RHIC}}, \href{https://doi.org/10.1140/epjc/s10052-011-1694-2}{\emph{Eur.
  Phys. J.} {\bfseries C71} (2011) 1694}
  [\href{https://arxiv.org/abs/1103.3511}{{\ttfamily 1103.3511}}].

\bibitem{Nara:2016hbg}
Y.~Nara, H.~Niemi, J.~Steinheimer and H.~St{\"o}cker, \emph{{Equation of state
  dependence of directed flow in a microscopic transport model}},
  \href{https://doi.org/10.1016/j.physletb.2017.02.020}{\emph{Phys. Lett.}
  {\bfseries B769} (2017) 543}
  [\href{https://arxiv.org/abs/1611.08023}{{\ttfamily 1611.08023}}].

\bibitem{Lambda-magmom}
L.~Schachinger, G.~Bunce, P.~Cox, T.~Devlin, J.~Dworkin et~al., \emph{{A
  Precise Measurement of the $\Lambda^0$ Magnetic Moment}},
  \href{https://doi.org/10.1103/PhysRevLett.41.1348}{\emph{Phys.Rev.Lett.}
  {\bfseries 41} (1978) 1348}.

\bibitem{Sigma-magmom}
C.~Ankenbrandt, J.~Berge, A.~Brenner, J.~Butler, K.~Doroba et~al., \emph{{A
  Precise Measurement of the Sigma+ Magnetic Moment}},
  \href{https://doi.org/10.1103/PhysRevLett.51.863}{\emph{Phys.Rev.Lett.}
  {\bfseries 51} (1983) 863}.

\bibitem{Adamczyk:2014vca}
{\scshape STAR} collaboration, \emph{{$\Lambda\Lambda$ Correlation Function in
  Au+Au collisions at $\sqrt{s_{NN}}=$ 200 GeV}},
  \href{https://doi.org/10.1103/PhysRevLett.114.022301}{\emph{Phys. Rev. Lett.}
  {\bfseries 114} (2015) 022301}
  [\href{https://arxiv.org/abs/1408.4360}{{\ttfamily 1408.4360}}].

\bibitem{Gongyo:2017fjb}
S.~Gongyo et~al., \emph{{Most Strange Dibaryon from Lattice QCD}},
  \href{https://doi.org/10.1103/PhysRevLett.120.212001}{\emph{Phys. Rev. Lett.}
  {\bfseries 120} (2018) 212001}
  [\href{https://arxiv.org/abs/1709.00654}{{\ttfamily 1709.00654}}].

\bibitem{Adare:2009qk}
{\scshape PHENIX} collaboration, \emph{{Detailed measurement of the $e^+ e^-$
  pair continuum in $p+p$ and Au+Au collisions at $\sqrt{s_{NN}} = 200$ GeV and
  implications for direct photon production}},
  \href{https://doi.org/10.1103/PhysRevC.81.034911}{\emph{Phys. Rev.}
  {\bfseries C81} (2010) 034911}
  [\href{https://arxiv.org/abs/0912.0244}{{\ttfamily 0912.0244}}].

\bibitem{Agostinelli:2002hh}
{\scshape GEANT4} collaboration, \emph{{GEANT4: A Simulation toolkit}},
  \href{https://doi.org/10.1016/S0168-9002(03)01368-8}{\emph{Nucl. Instrum.
  Meth.} {\bfseries A506} (2003) 250}.

\bibitem{E16}
S.~Yokkaichi et~al.{\emph{Proposal for J-PARC E16} (2006) }.

\bibitem{jparcweb:2016}
{\emph{http://asrc.jaea.go.jp/soshiki/hadron/jparc-hi} }.

\end{thebibliography}\endgroup

%\begin{thebibliography}{99}
%\bibitem{...}
%....

%\end{thebibliography}

\end{document}